\documentclass[a4paper,12pt]{article}
\usepackage[cp1251]{inputenc}
\usepackage{graphics}
\usepackage{amssymb,amsmath,latexsym}
\usepackage[english]{babel}
\usepackage{indentfirst}
\usepackage{graphicx,times}
\usepackage{natbib}
\begin{document}

\bigskip
\centerline {\bf The study of time series of monthly averaged
values of solar 10.7 cm radio flux from 1950 to 2010}
\bigskip

\centerline {E.A. Bruevich $^{a}$ , V.V. Bruevich $^{b}$, G.V.
Yakunina $^{c}$ }

\centerline {\it $^{a,b,c}$ Sternberg Astronomical Institute, Moscow
State
 University,}
\centerline {\it Universitetsky pr., 13, Moscow 119992, Russia}\

\centerline {\it e-mail:  $^a${red-field@yandex.ru},
$^b${brouev@sai.msu.ru}, $^c${yakunina@sai.msu.ru}}\

\bigskip
{\bf Abstract.} Prior to 1947, the activity of the Sun was assessed
by the relative numbers of sunspots (W). The 10.7 cm radio emission
(frequency of 2.8 GHz) for observations of the variability of
radiation of chromosphere and the lower corona 10.7 cm radio flux
($F_{10.7}$) became used from 1947. For the $F_{10.7}$ are available
more detailed observational archive data, so this activity index
more often than the other indices is used in the prediction and
monitoring of the solar activity. We have made the analysis of time
series of $F_{10.7}$ with the use of different mother wavelets:
Daubechies 10, Symlet 8, Meyer, Gauss 8 and Morlet. Wavelet spectrum
allows us not only to identify cycles, but analyze their change in
time. Each wavelet has its own characteristic features, so sometimes
with the help of different wavelets it can be better identify and
highlight the different properties of the analyzed signal. We
intended to choose the mother wavelet, which is more fully gives
information about the analyzed index $F_{10.7}$. We have received,
that all these wavelets show similar values to the maximums of the
cyclic activity. However, we can see the difference when using
different wavelets. There are also a number of periods, which,
perhaps, are the harmonics of main period. The mean value of 11-year
cycle is about 10.2 years.  All the above examples show that the
best results we get when using wavelets Morlet, Gauss (real-valued)
and multiparameter family of wavelets Morlet and Gauss
(complex-valued).

\bigskip
{\it Key words.}  Solar cycle: observations, solar activity
indices, wavelet spectrum.

\vskip12pt \centerline {\bf1. Introduction} \vskip12pt

The nature of solar activity is very complex. It has become of
great practical and societal importance to predict solar activity
and space climate. There are some important global indices of
solar activity which allow us to monitor the situation on the sun
and to build various forecasts. We have studied earlier these
indices and their mutual correlation during the solar cycles 21 -
23 in (Bruevich \& Yakunina 2011; Borisov {\it et al.} 2012). The
high degree of correlation of the 10.7 cm flux with all global
indices suggests some dependence upon common plasma parameters and
that their sources are spatially close. Another strong
correspondence is between 10.7 cm flux and full-disc X-ray flux.
When activity is high, they are well-correlated; however, when
activity is low, the X-rays are too weak to be detected, while
some 10.7 cm emission in excess of the "Quiet Sun Level" is always
present (Kruger 1979). Our study of the connection between 10.7 cm
flux and full-disc X-ray flux (Bruevich \& Yakunina 2011; Bruevich
\& Bruevich 2013) also confirm the conclusions of Kruger (1979).
Thus we have enhanced 10.7 cm radiation when the temperature,
density and magnetic fields are enhanced. So $F_{10.7}$ is a good
measure of general solar activity.

\vskip12pt \centerline {\bf2. 10.7 - cm solar radio flux and other
global activity indices} \vskip12pt

The most popular index - sunspot number SSN (also known as the
International sunspot number, relative sunspot number, or Wolf
number) is a quantity that measures the number of sunspots and
groups of sunspots present on the surface of the sun.

The historical sunspot record was first put by Wolf in 1850s and
has been continued later in the 20th century until today. Wolf's
original definition of the relative sunspot number for a given day
as $R = 10 \cdot$ Number of Groups + Number of Spots  visible on
the solar disk has stood the test of time. The factor of 10 has
also turned out to be a good choice as historically a group
contained on average ten spots. Almost all solar indices and solar
wind quantities show a close relationship with the SSN. (Svalgaard
{\it et al.} 2011; Svalgaard \& Cliver  2010). In our paper we use
the proper homogeneity calibrations of SSN from (National
Geophysical Data Center. Solar Data Service 2013), see Figure 1.

At the present time the 10.7 - cm solar radio flux $F_{10.7}$ is
measured at the Dominion Radio Astrophysical observatory in
Penticton, British Columbia  by the Solar Radio Monitoring
Programme. $F_{10.7}$  is a useful proxy for the combination of
chromospheric, transition region, and coronal solar EUV emissions
modulated by bright solar active regions whose energies at the Earth
are deposited in the thermosphere.(Tobiska {\it et al.} 2008)
pointed the high EUV - $F_{10.7}$ correlation and used this in the
Earth's atmospheric density models.

According to (Tapping \& DeTracey 1990) the 10.7 - cm emission from
the whole solar disc can be separated on the basis of characteristic
time-scales into 3 components: (i) transient events associated with
flare and similar activity having duration less than an hour; (ii)
slow variation in intensity over hours to years, following the
evolution of active regions in cyclic solar activity designated as
S-component; (iii) a minimum level below which the intensity never
falls - the "Quiet Sun Level". The excellent correlation of
S-component at 10.7 cm wavelength with full-disc flux in Ca II and
MgII was discussed by (Donnelly {\it et al.} 1983 ). The 10.7 cm
flux resembles the integrated fluxes in UV and EUV well enough to be
used as their proxy (Chapman  \& Neupert 1974; Donnelly {\it et
al.}1983; Bruevich \& Nusinov 1984; Nicolet \& Bossy 1985; Lean
1987)

This radio emission comes from high part of the chromosphere and low
part of the corona. $F_{10.7}$ radio flux  has two different
sources: thermal bremsstrahlung (due to electrons radiating when
changing direction by being deflected by other charged participles -
free-free radiation) and gyro-radiation (due to electrons radiating
when changing direction by gyrating around magnetic fields lines).
The (iii) a minimum level component (when SSN is equal to zero as it
was at the minimum of the cycle 24 and local magnetic fields are
negligible) is defined by free-free source. When the local magnetic
fields become strong enough at the beginning of the rise phase of
solar cycle and solar spots appear the gyro-radiation source of
$F_{10.7}$ radio flux begins to prevail over free-free so (i) and
(ii) components begin to grow strongly.

The S-component comprises the integrated emission from all sources
on solar disc. It contains contribution from free-free and
gyroresonance processes, and perhaps some non-thermal emission
(Gaizauscas \& Tapping 1998). The relative magnitude of these
processes is also a function of observing wavelength. Observations
of emission from active regions over the wavelength range 21-2 cm
suggest that at 21 cm, free-free emission is dominant, whereas at 6
cm, the contribution from gyroresonance is larger. At a wavelength
of 10 cm, the two processes are roughly equal in importance. At a
wavelength of 2-3 cm, the emission is again mainly free-free,
possible with a non-thermal component (Gaizauscas \& Tapping 1998).
The spatial distributions of two thermal processes are different;
the gyroresonant emission originates chiefly in the vicinity of
sunspots, where the magnetic fields are strong enough, while the
free-free emission is more widely-distributed over the host region
complex (Tapping \& DeTracey 1990).

\begin{figure}[h!]
 \centerline{\includegraphics[width=80mm]{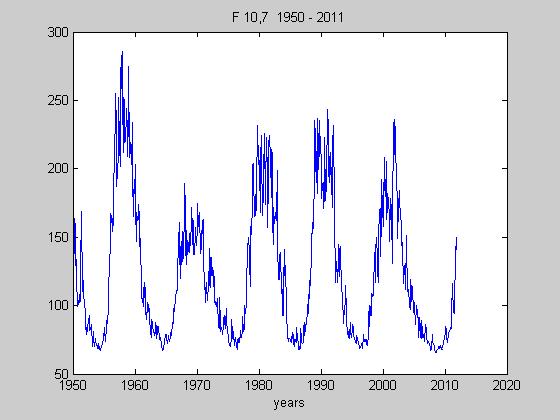}}
\caption{The time series of monthly average of 10.7 - cm solar
radio flux from 1950 to 2010. According to National Geophysical
Data Center of Solar and Terrestrial Physics data.}
\label{Fi:Fig1}
\end{figure}

The intensities of the Ca II and Mg II spectral lines are primary
functions of chromospheric density and temperature, while the soft
X-rays are produced in the corona. The high degree of correlation
of the 10.7 cm flux with all these quantities suggests some
dependence upon common plasma parameters and that their sources
are spatially close. Another strong correspondence is between 10.7
cm flux and full-disc X-ray flux. When activity is high, they are
well-correlated; however, when activity is low, the X-rays are too
weak to be detected, while some 10.7 cm emission in excess of the
"Quiet Sun Level" is always present (Kruger 1979). Our study of
the connection between 10.7 cm flux and full-disc X-ray flux
(Bruevich, \& Yakunina 2011) also confirm the conclusions of
(Kruger 1979).

 Thus we have enhanced 10.7 cm radiation when the
temperature, density and magnetic fields are enhanced. So $F_{10.7}$
is a good measure of general solar activity.

The 280 nm Mg II solar spectrum band contains photospheric
continuum and chromospheric line emissions. The Mg II {\it h} and
{\it k} lines at 279.56 and 280.27 nm, respectively, are
chromospheric in origin while the weakly varying wings or nearby
continuum are photospheric in origin.

Solar irradiance is the total amount of solar energy at a given
wavelength received at the top of the earth's atmosphere per unit
time. When integrated over all wavelengths, this quantity is
called the total solar irradiance (TSI) previously known as the
solar constant. Regular monitoring of TSI has been carried out
since 1978. From 1985 the total solar irradiance was observed by
Earth Radiation Budget Satellite (EBRS). We use the NGDC TSI data
set from combined observational data of several satellites which
were collected in NASA archive data (National Geophysical Data
Center. Solar Data Service 2013). The importance of UV/EUV
influence to TSI variability (Active Sun/Quiet Sun) was pointed by
(Krivova \& Solanki 2008).

\vskip12pt
\centerline {\bf3. The choice of mother wavelet for our
$F_{10.7}$ study}

The history of wavelets is not very old, at most 15 to 20 years.
There are lots of successes for the community to share. Fourier
techniques were liberated by the appearance of windowed Fourier
methods that operate locally on a time-frequency approach. The
wavelets bring their own strong benefits to that environment: a
local outlook, a multiscaled outlook, cooperation between scales,
and a time-scale analysis. They demonstrate that sines and cosines
are not the only useful functions and that other bases made of
weird functions serve to look at new signals, as strange as most
fractals or some transient signals. The choice of wavelet is
dictated by the signal or image characteristics and the nature of
the signal application. If you understand the properties of the
analysis and synthesis wavelet, you can choose a wavelet that is
optimized for your application.

We tried to choose the wavelet most useful for the analysis of
observational data of different indices of solar activity. In this
paper we analyzed the different mother wavelets for the study of
$F_{10.7}$ data (as a measure of general solar activity).

Modern methods of spectral analysis, in particular wavelet
analysis, allow us to successfully carry out the processing of
data of observations of the solar activity on different time
scales (Morozova {\it et al.} 1999).

In this paper we have made the analysis of time series of $F_{10.7}$
with the use of different mother wavelets: Daubechies 10, Simlet 8,
Meyer, Gauss 8 and Morlet (real and complex). It's known that
Fourier analysis consists of breaking up a signal into sine waves of
various frequencies.  Similarly, wavelet analysis is the breaking up
of a signal into shifted and scaled versions of the original (or
mother) wavelet.

The wavelets bring their own strong benefits to that environment:
a local outlook, a multiscaled outlook, cooperation between
scales, and a time-scale analysis.  They demonstrate that sines
and cosines are not the only useful functions and that other bases
made of weird functions serve to look at new foreign signals, as
strange as most fractals or some transient signals. The wavelets
are the localized functions constructed with help of one so-called
mother wavelet $\psi (t)$  by shift operation on argument (b) :

    $$ \psi_{ab}(t)  = (1/ \sqrt{ \left| a \right| } \cdot  \psi ((t-b) / a)               $$

and scale change (a):

$$      \psi((t-b)/a)                                            $$

The wavelet  time-scale spectrum C(a,b)  is the two-arguments
function. Note than scale change "a"  is measured in
reversed-frequency units  and argument "b" is measured in time
units:

$$      C(a,b) =   (1/ \sqrt{ \left| a \right| } \int_{-\infty}^{\infty} S(t) \cdot   \psi ((t-b) / a) dt                  $$

Wavelet analysis is successfully applied for the processing of
time series of astronomical observations. In (Vityazev 2001) it
has been analyzed the possibility of using of various mother
wavelets for a set of astronomical applications. The choice of
wavelet is dictated by the signal or image characteristics and the
nature of the application. If you understand the properties of the
analysis and synthesis wavelet, you can choose a wavelet that is
optimized for your application.  Wavelet families vary in terms of
several important properties.  Examples include:

 - support of the wavelet in time and frequency and rate of decay;

 - symmetry or antisymmetry of the wavelet. The accompanying perfect
reconstruction filters have linear phase;

- number of vanishing moments. Wavelets with increasing numbers of
vanishing moments result in sparse representations for a large class
of signals and images;

- regularity of the wavelet.  Smoother wavelets provide sharper
frequency resolution. Additionally, iterative algorithms for wavelet
construction converge faster.

\vskip12pt \centerline {\bf3.1 Morlet wavelet }

The Morlet wavelet is suitable for continuous analysis. The Morlet
wavelet - is the plane wave, modulated by Gaussian function:
 $$   \psi(t)=~e^{-\frac{t^2}{a^2}} \cdot  e^{i 2\pi t}            $$

On Figure 2 we see the Morlet wavelet function $\psi$.

\begin{figure}[h!]
 \centerline{\includegraphics[width=60mm]{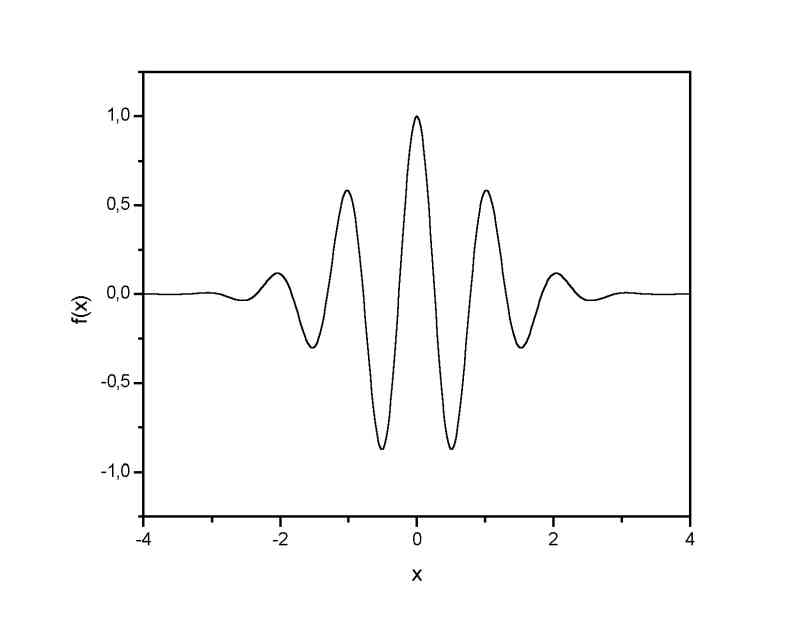}}
\caption{The Morlet wavelet function $\psi$ .} \label{Fi:Fig2a}
\end{figure}

In mathematics, the Morlet wavelet (or Gabor wavelet) is a wavelet
composed of a complex exponential (carrier) multiplied by a
Gaussian window (envelope). In 1946, physicist Dennis Gabor,
applying ideas from quantum physics, introduced the use of
Gaussian-windowed sinusoids for time-frequency decomposition,
which he referred to as atoms, and which provide the best
trade-off between spatial and frequency resolution. These are used
in the Gabor transform, a type of short-time Fourier transform. In
1984, Jean Morlet introduced Gabor's work to the seismology
community and, with Goupillaud and Grossmann, modified it to keep
the same wavelet shape over equal octave intervals, resulting in
the first formalization of the continuous wavelet transform.

On Figure 3 we see the results of the continuous wavelet transform
analysis (with help of Morlet mother vawelet) of time series of
monthly averaged $F_{10.7}$. Plane XY corresponds to the
time-frequency plane (a, b): a - Y (Cyclicity, years), b - X-
(Time, years). The C(a,b) coefficients characterizing the
probability amplitude of regular cyclic component localization
exactly at the point (a, b), are laid along the Z axis. At Figure
3 we see the projection of C(a,b) to (a, b) or (X, Y) plane. This
projection on the plane (a, b) with isolines allows to trace the
changes of the coefficients on various scales in time and reveal a
picture of local extremum of these surfaces. It is the so-called
skeleton of the structure of the analyzed process. In (Vityazev
2001) for processing of time series of astronomical observations
the preference is given to the Morlet mother wavelet. The
interpretation of Morlet-wavelet images is similar to the
interpretation of the results of Fourier analysis of data sets. We
can also note that the configuration of Morlet wavelet is very
compact in frequency, which allows us the most accurately
(compared with other wavelets) to determine the localization of
instantaneous frequency of observed signal. We can see the main
11-yr cycle of activity. The most probable value of this cyclicity
is about 10 years. We also can see a set of quassi-biennial cycles
inside of every 11-yr cycle which duration vary from 3 to 2.5
year.

\begin{figure}[h!]
 \centerline{\includegraphics[width=90mm]{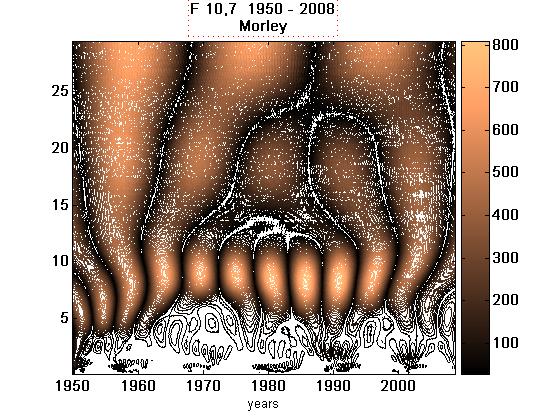}}
\caption{The analysis of time series of monthly averaged $F_{10.7}$
with the use of Morlet mother wavelet.} \label{Fi:Fig2b}
\end{figure}

\vskip12pt \centerline {\bf3.2 Daubechies wavelet} \vskip12pt

We see the Daubechies 8 wavelet function $\psi$ on Figure 4.

\begin{figure}[h!]
 \centerline{\includegraphics[width=60mm]{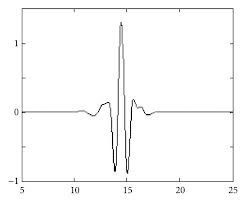}}
\caption{The Daubechies 8 wavelet function $\psi$ .}
\label{Fi:Fig3a}
\end{figure}

The Daubechies wavelets, based on the work of Ingrid Daubechies,
are a family of orthogonal wavelets defining a discrete wavelet
transform and characterized by a maximal number of vanishing
moments for some given support. With each wavelet type of this
class, there is a scaling function (called the father wavelet)
which generates an orthogonal multiresolution analysis.

On Figure 5 we demonstrate our analysis of radio emission $F_{10.7}$
with the help of Daubechies 8 mother wavelet. This study of time
series of $F_{10.7}$ shows that the previous cycles affect the
subsequent cycles. This is connected with peculiarity of this
wavelet, its wider coverage of the sample studied observations. But
such a wide filter leads to more blurred values which determine the
maximum probability of the determination of duration of the cycle.

\begin{figure}[h!]
 \centerline{\includegraphics[width=90mm]{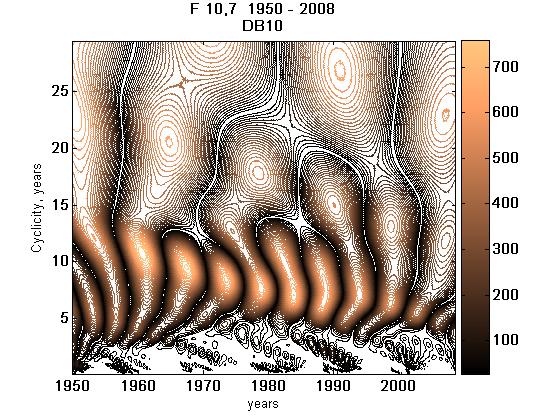}}
\caption{The analysis of time series of monthly averaged $F_{10.7}$
with the use of Daubechies 8 mother wavelet.} \label{Fi:Fig3b}
\end{figure}

\vskip12pt \centerline {\bf3.3 Simlet wavelet} \vskip12pt

The Simlet N wavelets are also known as Daubechies'
least-asymmetric wavelets. The symlets are more symmetric than the
extremal phase wavelets. In Simlet N, N is the number of vanishing
moments.

On Figure 6 we demonstrate the analysis of radio emission $F_{10.7}$
with the help of Simlet 8 mother wavelet. We can show that
time-frequency parameters in this case have much more blurred
contours around the maximums. Thus, errors in determining the most
probable values of the cycle's duration are increased compared with
the study of a given series of observations using the wavelet
Morlet.

\begin{figure}[h!]
 \centerline{\includegraphics[width=90mm]{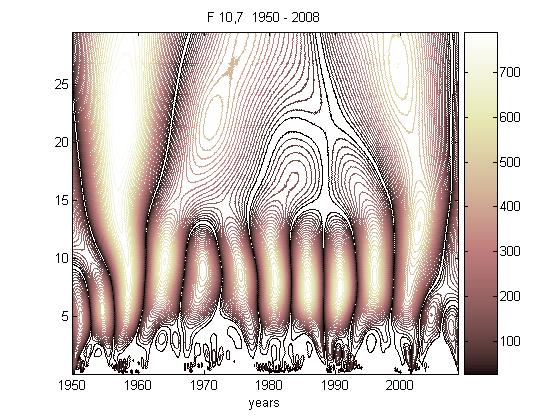}}
\caption{The analysis of time series of monthly averaged $F_{10.7}$
with the use of Simlet 8 mother wavelet.} \label{Fi:Fig4b}
\end{figure}

\vskip12pt \centerline {\bf3.4 Meyer wavelet}

\begin{figure}[h!]
 \centerline{\includegraphics[width=80mm]{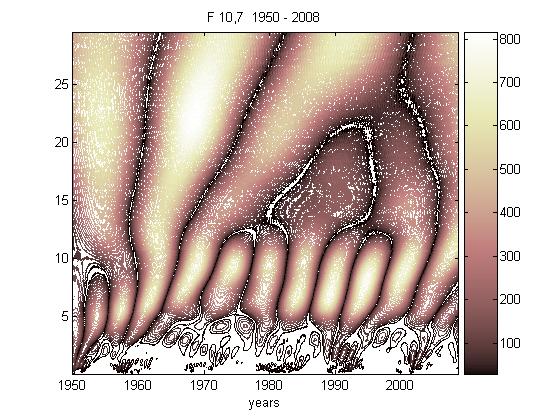}}
\caption{The analysis of time series of monthly averaged $F_{10.7}$
with the use of Meyer mother wavelet.} \label{Fi:Fig5b}
\end{figure}

The Meyer's wavelets are the orthogonal wavelets proposed by Yves
Meyer. These mother wavelets are defined in a such a way to avoid
the slow decay in the space domain.

On Figure 7 we demonstrate the analysis of radio emission $F_{10.7}$
with the help of Meyer mother wavelet. We also see that
time-frequency parameters in this case are not as good as in the
case of Morlet wavelet. Thus, errors in cycle's duration
determination are increased compared with the study of a given
series of observations using the Morlet wavelet.

\vskip12pt \centerline {\bf3.5 Gaussian wavelet}

\begin{figure}[h!]
 \centerline{\includegraphics[width=80mm]{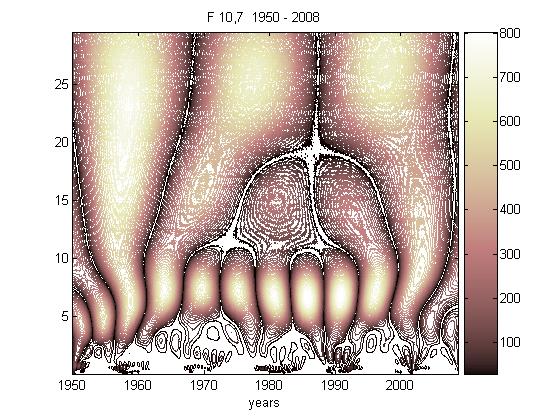}}
\caption{The analysis of time series of monthly averaged
$F_{10.7}$ with the use of Gaussian 8 mother wavelet.}
\label{Fi:Fig6b}
\end{figure}

The results of processing observation series using the Gaussian
wavelet are very similar to the results of processing with the
Morlet wavelet. We also can see along with a basic 11-yr cycle of
activity the quassi-biennial cyclicity.

On Figure 8 we demonstrate the analysis of radio emission $F_{10.7}$
with the help of Gaussian 8 mother wavelet. We see that
time-frequency parameters in this case are practically coinciding
with the characteristics obtained with the use of the mother wavelet
Morlet. Errors in determination of the most probable values of
duration of the cycles are not more than in case when we use the
Morlet wavelet. The differences between these wavelet studies we see
in small details, more concerning quassi-biennial cycles.

\vskip12pt \vskip12pt \centerline {\bf3.6 Complex Morlet  wavelet}
\vskip12pt \vskip12pt

A complex Morlet wavelet is defined by:

 $$   \psi(x)=~\frac{1}{\sqrt{\pi f_b}}\cdot  e^{ 2i\pi f_c x} \cdot e^{-\frac{~x^2}{f_b}}           $$

A complex Morlet wavelet is depending on two parameters:  $f_b$ is a
bandwidth parameter and $f_c$ is a wavelet center frequency.

\bigskip
\begin{figure}[h!]
 \centerline{\includegraphics[width=90mm]{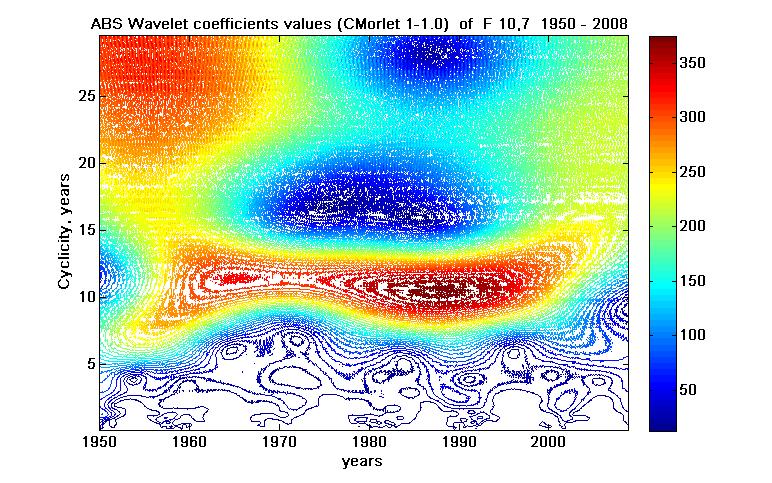}}
\caption{The time series of $F_{10.7}$ (1975 - 2010 yr) and
analysis of the F10.7 data with the use of complex Morlet 1-1.0
wavelet. } \label{Fi:Fig7}
\end{figure}

On Figure 9 we show the analysis of time series of monthly averaged
$F_{10.7}$ (1974 - 2010 years) with help a complex Morlet wavelet.
In this case, there are two additional parameters that can be varied
in accordance with the objectives set tasks and, as in the case of
Fourier analysis, we get the array of coefficients which give us
information not only about the frequency-temporal distribution of
the amplitude and about the frequency-temporal distribution of the
phase of the signal. The parameters of the mother wavelet
Complex-valued Morlet 1-1.0 have the best determination of the
11-year cycle unless of cycles not well pronounced and irregular. It
is also seen that the magnitude of the 23-rd cycle of activity for
more than 12 years (this value we do not get with use of real-valued
wavelet Morlet as a result analysis).

\vskip12pt \centerline {\bf4. Summary and conclusions } \vskip12pt

Wavelet spectrum allows us not only to identify cycles, but analyze
their change in time. Each wavelet has its own characteristic
features, so sometimes with the help of different wavelets it can be
better identify and highlight the different properties of the
analyzed signal. We intended to choose the mother wavelet, which is
more fully gives information about the analyzed solar index
$F_{10.7}$. These are Morlet and Gauss real-valued and Morlet
complex-valued wavelets. With these wavelets we can study the solar
cyclicity evolution of the most accurate form in every moment of
time. We can see also a number of periods, which, perhaps, are the
harmonics of main period. The complex-valued Morlet and Gauss
wavelet analysis gives us the additional information about signal
phase evolution. Note than the Daubechies 10 wavelet-analysis (more
wide filter window) allows us to analyze the influence of the
previous cycle to the next. The Mexican hat wavelet inhibits the
main cyclicity and allows us to analyze the cyclicity of second-
order. The analysis with all mother wavelet shows that the mean
value of 11-year cycle is about 10.2 years during the period 1950 -
2000. The complex-valued Morlet wavelet analyzes shows the more long
duration of 11-yr cyclicity for the cycle 23 - about 12 yr.

\bigskip
{\bf Acknowledgements} The authors thank the RFBR grant
11-02-00843ap for support of the work.

\bigskip
{\bf References}
\bigskip

Alekseev, I.Yu. \& Gershberg, R.E. 1996, {\it On spotting of red
dwarf stars: direct and inverse problem of the construction of zonal
model, Astronomy Report}, {\bf 73}, 589.

Baliunas, S.L., \& Donahue, R.A., \& Soon, W.H. et al. 1995, {\it
Chromospheric variations in main-sequence stars, Astrophysical
Journal}, {\bf  438}, 269.

Borisov, A.A. \& Bruevich, E.A. \& Rozgacheva, I.K. \& Yakunina,
G.V. 2012, {\it Solar Activity Indices in the Cycles 21-23, The Sun:
New Challenges, Astrophysics and Space Science Proceedings, Vol-ume
30. ISBN 978-3-642-29416-7. Spring-er-Verlag Berlin Heidelberg},
221.

Bowman, B.R, \&  Tobiska, W.K., \& Marcos, F.A. et al., 2008, {\it A
New Empirical Thermospheric Density Model JB2008 Using Solar and
Geomagnetic Indices, AIAA/AAS Astrodynamics Specialist Conference,
AIAA 2008-6438}.

Bruevich, E.A., \& Alekseev I.Yu. 2007, {\it Spotting in stars with
a low level of activity, close to solar activity. Astrophysics},
{\bf 50}, No 2, 187.

Bruevich, E.A., \& Bruevich, V.V. 2013, {\it Changed relation
between solar 10.7 cm radio flux and some activity indices
 which describe the radiation at different
altitudes of atmosphere. ArXiv e-prints}, arXiv:1304.4545v1.

Bruevich, E.A., \& Kononovich E.V. 2011, {\it Solar and Solar-type
Stars Atmosphere's Activity at 11-year and Quasi-biennial Time
Scales. Vestn. Mosk. Univ. Fiz. Astron., {\bf  N1}, 70. ArXiv
e-prints}, arXiv:1102.3976v1.

Bruevich, E.A., \& Nusinov A.A, 1984, {\it Spectrum of short-wave
emission for aeronomical calculations for different levels of solar
activity, Geomagnetizm i Aeronomiia}, {\bf24}, 581.

Bruevich, E.A., \& Yakunina, G.V., 2011 {\it Solar Activity Indices
in the Cycles 21 - 23, arXiv:1102.5502v1}

Chapman, R.D., \& Neupert, W.M., 1974,  {\it Slowly varying
component of extreme ultraviolet solar radiation and its relation to
solar radio radiation,  J. Geophys. Res.}, {\bf 79}, 4138.

Donnelly, R.F., \& Heath, D.F., \& Lean, J. L. \& Rottman, G.J.,
1983, {\it Differences in the temporal variations of solar UV flux,
10.7-cm solar radio flux, sunspot number, and Ca-K plage data caused
by solar rotation and active region evolution, J. Geophys. Res.},
{\bf88}, 9883.

Fligge, M., \& Solanki, S.K., \& Unruh, Y.C., \& Frohlich, C., \&
Wehrli, C. 1998,  {\it A model of solar total and spectral
irradiance variations. Astronomy \& Astrophys.} {\bf 335}, 709.

Floyd, L., \& Newmark, J., \& Cook, J., \& Herring, L., \& McMullin,
D. 2005, {\it Solar EUV and UV spectral irradiances and solar
indices. Journal of Atmospheric and Solar-Terrestrial Physics}, {\bf
67}, 3.

Gaizauscas, V. \& Tapping, K.F., 1988, {\it Compact sites at 2.8 cm
wavelength  of microwave emission  inside solar active regions.
Astrophys. J.}, {\bf 325}, 912.

Ishkov, V.N. 2009, {\it 1st Workshop on the activity cycles on the
Sun and stars, Moscow, 18-10 December, Edited by EAAO,
St-Petersburg}, 57.

Janardhan, P. \&  Susanta, K.B. \& Gosain, S., 2010 {\it Solar Polar
Fields During Cycles 21 - 23: Correlation with Meridional Flows,
Solar Physics}, {\bf 267}, 267.

Janardhan, P. \& Bisoi, S.K. \& Ananthakrishnan, S \& Tokumaru, M \&
Fujiki, K. 2011 {\it Interplanetary scintillation signatures in the
inner heliosphere, Geophysical Research Letters}, {\bf 38}, Issue
20, L20108.

Kleczek, J. 1952, {\it Catalogue de l'activite' des e'ruptions
chromosphe'riques. Publ. Inst. Centr. Astron.}, {\bf 22}.

Krivova, N.A., \& Solanki, S.K., \& Fligge, M., \& Unruh, Y. C.
2003,  {\it Reconstruction of solar total and spectral irradiance
variations in cycle 23: is solar surface magnetism the cause?,
Astron. Astrophys.} {\bf 339}, L1.

Krivova, N. A., \& Solanki, S. K. 2008, {\it Models of solar
irradiance variations: current status. Journal of Astrophysics and
Astronomy}, {\bf 29}, 151.

Kruger, A. 1979, {\it Introduction to Solar Radio Astronomy and
Radio physics, D. Reidel Publ. Co., Dordrecht, Holland}.

Lean, J. L., 1990, {\it A comparison of models of the Sun's extreme
ultraviolet irradiance variations, J. Geophys. Res.}, {\bf95},
11933.

Livingston, W., \& Penn, M. J.,  \& Svalgaard  L., 2012, {\it
Decreasing Sunspot Magnetic Fields Explain Unique 10.7 cm Radio
Flux, Astrophys. J.}, {\bf 757}, N1, L8.

Lukyanova, R., \& Mursula, K. 2011, {\it Changed relation between
sunspot numbers, solar UV/EUV radiation and TSI during the declining
phase of solar cycle 23.
 Journal of Atmospheric and Solar-Terrestrial Physics}
{\bf 73}, 235.

Nagovitsyn, Y.A., \& Pevtsov, A.A., \& Livingston W.C. 2012, {\it On
a possible explanation of the long-term decrease in sunspot field
strength, Astrophysical Journal Letters}, {\bf 758}, L20.

National Geophysical Data Center. Solar-Geophysical Data Reports. 54
Years of Space Weather Data. 2009, {\it
http://www.ngdc.noaa.gov/stp/solar/sgd.html}.

National Geophysical Data Center. Solar Data Service. Sun, solar
activity and upper atmosphere data. 2013, {\it
http://www.ngdc.noaa.gov/stp/solar/solardataservices.html}.

Nicolet, M., \& Bossy, L., 1985, {\it Solar Radio Fluxes as indices
of solar activity, Planetary Space Sci.}, {\bf33}, 507.

Penn, M.J., \& Livingston, W.C. 2006, {\it Temporal Changes in
Sunspot Umbral Magnetic Fields and Temperatures, Astrophysical
Journal Letters}, {\bf649}, L45.

Pevtsov A. A., \& Nagovitsyn, Y. A., \& Tlatov, A. G., \& Rybak, A.
L. 2011 {\it Long-term Trends in Sunspot Magnetic Fields,
Astrophysical Journal Letters}, {\bf742}, L36.

Skupin, J., \& Noyel, S.,\& Wuttke, M.W., \& Gottwald, M., \&
Bovensmann, H., \& Weber, M., \& Burrows, J. P. 2005, {\it SCIAMACHY
solar irradiance observation in the spectral range from 240 to 2380
nm, Advance Space Res.}, {\bf 35}, 370.

Svalgaard, L., \& Lockwood M., \& Beer J. 2011, {\it Long-term
reconstruction of Solar and Solar Wind Parameters}, {\it
http://www.leif.org/research/Svalgaard\_ISSI\_Proposal\_Base.pdf}.

Svalgaard, L. \& Cliver E.W. 2010, {\it Heliospheric magnetic field
1835-2009, J. Geophys. Res.}, {\bf115}, A09111.

Tapping, K.F., \& DeTracey, B., 1990, {\it The origin of the 10.7 cm
flux, Solar Physics}, {\bf127}, 321.

Tobiska, W.K., \&  Bouwer S.D., \& Bowman, B.R., {\it The
development of new solar indices for use in thermospheric density
modeling, 2008, J. Atmospheric \& Solar-Terrestrial Phys.},{\bf 70},
803.

Viereck, R., \& Puga, L., \& McMullin, D., \& Judge, D., \& Weber,
M. \& Tobiska, K. 2001, {\it The MgII index: a proxy for solar EUV.
Journal of Geophysical Research} , {\bf 73}, No 7, 1343.

Viereck, R.A., Floyd, L.E., Crane, P.C., Woods, T.N., Knapp, B.G.,
Rottman, G., Weber, M., Puga, L.C. and Deland, M.T. 2004, {\it A
composite MgII index spanning from 1978 to 2003. Space Weather},
{\bf 2}, No. 10, doi:10.1029/2004SW000084.

Vitinsky, Yu.,\& Kopezky, M., \& Kuklin G., 1986. {\it The sunspot
solar activity statistik}, Moscow,  Nauka.

\end{document}